\documentstyle[mnextra]{mn}
\def\r{{\bf r}}
\def\s{{\bf s}}
\def\v{{\bf v}}
\def\x{{\bf x}}

\def\k{{\bf k}}

\def\rhobar{{\bar \rho}}
\def\etal{et al.\ }
\renewcommand\d[1]{\,d\!#1}
\newcommand\dthre[1]{\,d^{3}\!#1}
\def\dr{\dthre{\r}}

\input{epsf}

\def\mod#1{\vert #1 \vert}
\def\avg#1{\langle #1 \rangle}

\def\eqnref#1{\hbox{(\ref{eqn:#1})}}
\def\figref#1{\hbox{Fig. \ref{fig:#1}}}
\def\secref#1{\hbox{Section \ref{sec:#1}}}
\def\tabref#1{\hbox{Table \ref{tab:#1}}}
\def\nbody{$N$-body\ }
\def\lsim{\mathrel{\hbox{\rlap{\hbox{\lower4pt\hbox{$\sim$}}}\hbox{$<$}}}}
\def\gsim{\mathrel{\hbox{\rlap{\hbox{\lower4pt\hbox{$\sim$}}}\hbox{$>$}}}}
\def\hmpc{h^{-1}\,{\rm Mpc}}
\def\hkpc{h^{-1}\,{\rm kpc}}
\def\himpc{h\,{\rm Mpc^{-1}}}
\def\kms{{\rm km\,s^{-1}}}
\nobrackets

\seceqnum      
\begin{document}

\title[Redshift-space distortion]{Modelling the redshift-space distortion of galaxy clustering}
\author[S.J. Hatton and S. Cole]{Steve Hatton\thanks{S.J.Hatton@durham.ac.uk} and Shaun Cole\thanks{Shaun.Cole@durham.ac.uk}\\
Department of Physics, University of Durham, Science
Laboratories, South Rd, Durham DH1 3LE.
}

\maketitle

\begin{abstract}
We use a set of large, high-resolution cosmological \nbody simulations
to examine the redshift-space distortions of galaxy clustering on
scales of order $10-200 h^{-1}$Mpc.  Galaxy redshift surveys currently in
progress will, on completion, allow us to measure the quadrupole
distortion in the 2-point correlation function, $\xi(\sigma,\pi)$, or
its Fourier transform, the power spectrum, $P(k,\mu)$, to a high
degree of accuracy.  On these scales we typically find a positive
quadrupole, as expected for coherent infall onto overdense regions and
outflow from underdense regions, but the distortion is substantially
weaker than that predicted by pure linear theory.  We assess two
models that may be regarded as refinements to linear theory, the
Zel'dovich approximation and a dispersion model in which the
non-linear velocities generated by the formation of virialized groups
and clusters are treated as random perturbations to the velocities
predicted by linear theory. We find that neither provides an adequate
physical description of the clustering pattern.  If used to model
redshift space distortions on scales for $10<\lambda <200 h^{-1}$Mpc
the estimated value of $\beta$ ($\beta=f(\Omega_0)/b$ where
$f(\Omega_0) \approx \Omega_0^{0.6}$ and $b$ is the galaxy bias
parameter) is liable to systematic errors of order ten per cent or more.  We 
discuss how such systematics can be avoided by i) development of
a more complete model of redshift distortions and ii) the
direct use of galaxy catalogues generated from non-linear
\nbody simulations.
\end{abstract}



\section{Introduction}
\label{sec:intro}

In early galaxy redshift surveys (eg. \cite{jackson,greg}), some of the
most striking artefacts observed were the so-called `fingers of God',
ridges in the galaxy distribution pointing directly at the observer.
What had been observed was the first evidence of the redshift-space
distortion of galaxy clustering. If galaxy motions were perfectly
described by pure Hubble flow, then a galaxy's redshift would be an
accurate indicator of its distance and the clustering pattern observed
in galaxy redshift surveys would be statistically isotropic.  The
observed anisotropic clustering pattern is generated by galaxy
peculiar velocities which perturb galaxy redshifts and hence their
inferred distances.  On small scales, the non-linear velocities of
virialized groups and clusters create the `fingers of God' by
stretching out these structures along the line of sight.  On large
scales the clustering pattern is predicted (as suggested by 
Sargent \& Turner \shortcite{st77}) to be
compressed along the line of sight by coherent infall onto galaxy
clusters and super-clusters and outflow from voids and other
underdense regions.  Thus the anisotropy of galaxy clustering encodes
information about the galaxy velocity field and hence the underlying
mass density field that gave rise to it.  Measurements 
of redshift-space distortions can therefore be used to place
constraints on the density parameter, $\Omega_0$, and the bias parameter,
$b$, which relates the galaxy distribution to the mass distribution.

Interesting constraints on the combination $\beta \equiv
\Omega_0^{0.6}/b$ 
have been obtained by analysis of optical redshift
surveys (\cite{s-apmIII,dukst})  and IRAS surveys (eg. 
\cite{ham93,FisherIRAS,CFW95} and more recent papers such 
as \cite{FN96,bwz97}).  However, the
statistical errors remain large and no consensus on the value of $\beta$
has been reached (for a review see \cite{SW96}).  The
next generation of large surveys, including the 2-degree-Field 
(2dF, \cite{colless96}) and Sloan Digital Sky Survey (SDSS, \cite{gw95}), 
will enable redshift-space distortions to be
measured with unprecedented accuracy.  These surveys should allow
$\beta$ to be measured to an accuracy of a few per cent, and possibly
for $\Omega_0$ and $b$ to be separately constrained.  To achieve this
accuracy and avoid systematic errors it is important that the
theoretical modelling of redshift-space distortions be accurate. 
This is a non-trivial requirement, as these surveys will provide 
their most precise measurements of the distortions on scales of 
$30$--$100
\hmpc$, where the predictions of pure linear theory are not expected to hold.

In this paper we use a set of high-resolution \nbody simulations 
which accurately follow the evolution of clustering over the range 
of scales which will be probed by the next-generation surveys.  
Rather than constructing mock galaxy catalogues we instead use the 
full simulations to quantify the redshift-space distortions 
as accurately as possible, and through this assess
the accuracy of the linear theory model and two proposed extensions
to it. 
In \secref{anis} we define our notation convention and introduce 
the quadrupole statistic which we use to assess the level of 
redshift-space distortion.  \secref{sims} details the parameters 
of our set of \nbody simulations, 
and explains the methods we employ to construct biased galaxy samples.
The linear theory prediction and the two other models
are described in \secref{mods}.
In \secref{anal} we compare the analytic models
with measurements of the redshift-space distortions in the 
\nbody simulations.
We discuss results in \secref{conc}.


\section{Anisotropy in redshift-space}
\label{sec:anis}
\subsection{The power spectrum}
Our method of analysis uses the power spectrum, $P(\k)$, the Fourier 
transform of the correlation function,
\begin{equation}
P(\k) = \int{ \xi(\r) {\rm{e}}^{i\k.\r} \dr} ,
\end{equation}
where we use the Fourier convention that the wavenumber, $k$, 
is the angular frequency corresponding to wavelength $\lambda$, ie. 
$k=2\pi/\lambda$.  The correlation function is given by
\begin{equation}
\xi(\r) = \avg{\delta(\x +\r) \delta(\x)},
\end{equation}
where $\delta(\r)$ is the fractional overdensity,
\begin{equation}
\delta(\r) = \frac{\rho(\r) - \rhobar}{\rhobar}.  
\end{equation}
The power spectrum has been used by many authors to study the growth 
of gravitational clustering in simulations, and to quantify the  
clustering observed in redshift surveys (for recent examples, see 
\cite{cfa2,tadros,lascamp}).  

When galaxy distances are measured in redshift space, their 
peculiar velocities (ie. relative to the pure
Hubble flow) distort the pattern of galaxy clustering by displacing
galaxy positions along the line of sight. Thus a galaxy whose true
position is $\r$ appears in redshift-space at the  position
\begin{equation}
\s= \r + U(\r) \hat \r,
\end{equation}
where the line-of-sight peculiar velocity $U(\r)= \v(\r).\hat\r$.
Here we have adopted units in which the Hubble constant has unit value,
we measure distance in units of $\hmpc$,
where $h\equiv H_0/(100\kms{\rm Mpc^{-1}})$,
and velocities in units of $100\kms$. 

In redshift space this displacement in a preferred direction causes 
the observed power spectrum to be  anisotropic, with 
different values for wave-vectors along the line of sight to those 
perpendicular to it.  
Thus, the redshift-space power spectrum can be thought 
of as a function of $k\, (= \mod{\k})$ and $\mu$, the cosine of the
angle between  wave-vector $\k$ and the line of sight,
\begin{equation}
P(\k) = P(k,\mu) .
\end{equation}
In what follows we adopt the convention that, by $P(\k)$ or $P(k,\mu)$ 
we refer to the redshift-space power spectrum, since this quantity depends 
on both the magnitude and direction of $\k$, whereas $P(k)$ 
represents the real space power spectrum, depending only on the scalar $k$. 

\subsection{The quadrupole ratio}

The redshift-space distortions can be conveniently quantified by 
a simple statistic.  The anisotropy in $P(\k)$ is symmetric in 
$\mu$, ie. $P(k,\mu) = P(k,-\mu)$, so the distortion
depends only on even powers of $\mu$.  To measure the extent of deformation 
from isotropy, we decompose the power spectrum into multipole moments using 
the Legendre polynomials, $L_{l}(\mu)$.  Thus,
\begin{equation}
P(k,\mu) = \sum_{l=0}^{\infty} P_{l}(k) L_{l}(\mu) ,
\end{equation}
where the sum is over even values of $l$.
From the orthogonality relation for Legendre Polynomials,
the multipole moments $P_l(k)$ can be found by evaluating
\begin{equation}
\label{eqn:multipol}
P_{l}(k) = \frac{2l+1}{2} \int_{-1}^{+1} P(k,\mu)L_{l}(\mu) \d{\mu} .
\end{equation}
We find that estimates of $P_{l}(k)$ rapidly become noisy for multipoles  
with $l > 2$, so we choose to use as the key statistic
for our analysis the quadrupole to monopole ratio, $P_{2}(k)/P_{0}(k)$.

We have chosen to carry out all our analysis in $k$-space using power
spectrum multipoles $P_l(k)$. However, our results can readily be translated
into predictions of the multipole moments of the correlation function,
$\xi_l(r)$ using the identities derived in appendix~B of Cole, Fisher
\& Weinberg (\shortcite{CFW94}).

\section{Simulations}
\label{sec:sims}

\begin{figure}
\centering
\centerline{\epsfxsize = 8.0 cm \epsfbox{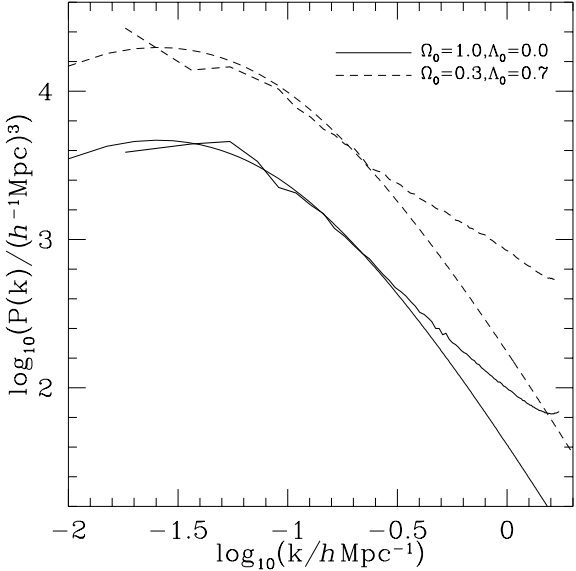}}
\caption{Power spectra of the real-space mass distributions of the
\nbody simulations described in \secref{sims}.  In each case the thin 
line is the linear theory, $\Gamma=0.25$ CDM spectrum \eqnref{p0} which 
was used to set up the initial conditions of the simulations, the thicker line 
is the evolved, non-linear power spectrum of the simulation measured at $z=0$.}
\label{fig:p0}
\end{figure}

In order to study the effects of non-linear gravitational evolution on 
redshift space clustering we examine a selection of \nbody simulations
with different values of the density parameter $\Omega_0$ and the galaxy bias.
The simulations we use are taken from a set of large volume simulations
which have been produced in order to make mock redshift catalogues 
of the Sloan Digital Sky Survey (\cite{gw95}) and the Anglo-Australian 
2-Degree Field Survey (\cite{colless96}), as will be described in  
\cite{mock}.  The full set of 
simulations consists of two series. The first is normalized to the amplitude
of CMB fluctuations in the COBE-DMR data set (\cite{smoot92})
and the second normalized to match present day observations of 
galaxy clusters and large scale structure. Both series cover a 
range in $\Omega_0$ and include both open models (with $\Lambda_0=0$),
and flat models ($\Omega_0+\Lambda_0=1$, where $\Lambda_0$
is the cosmological constant in units of $3H_0^2$).

The main cause of redshift-space distortion on small scales is
the random velocities of galaxies in groups and clusters. Since our aim is
to study deviations from linear theory on scales of $10$--$200 \hmpc$
this important non-linear effect most be included at a realistic
level. Therefore we have selected simulations from the series 
that are normalized to produce approximately the observed abundance 
of rich galaxy clusters. This set of simulations have the
rms mass fluctuation in spheres
of radius $8\hmpc$ set to $\sigma_8= 0.55 \Omega_0^{-0.6}$ (\cite{WEF93}).
The shape of the initial power spectrum is the same in each 
simulation and is given by the Bardeen et al. (\shortcite{BBKS}) formula for 
the cold dark matter spectrum,
\begin{equation}
P(k) \propto \frac{k \times [\ln(1+2.34q)/2.34q]^2}{[1+3.89q+(16.1q)^2+(5.46q)^3+(6.71q)^4]^{1/2}} .
\label{eqn:p0}
\end{equation}
Here $q=k/\Gamma$ and we set $\Gamma = 0.25$, as suggested
by observations of large scale structure (eg. \cite{FKP,apmIII}).
To illustrate our results we focus mainly on two models, the $\Omega_0=1$
model and a low density $\Omega_0=0.3$, $\Lambda_0=0.7$ model.

The initial particle positions and velocities were set by using the
Zel'dovich (\shortcite{ZA}) approximation to perturb particles from an
almost uniform ``glass'' distribution. This ``glass'' distribution
was produced using the technique described by White
(\shortcite{white94}), and Baugh, Gazta\~naga \& Efstathiou (\shortcite{baugh95}).  
The simulations were evolved using the AP$^3$M
code of Couchman (\shortcite{ap3m}) using $192^3$ particles in a
periodic cube of side $L_{\rm box}=345.6\hmpc$ (comoving distance).  
The softening
parameter of AP$^3$M's triangular-shaped cloud force law was set to
$\eta=270\hkpc$, $15$ per cent of the grid spacing.  This choice corresponds
approximately to a gravitational softening length
$\epsilon=\eta/3=90\hkpc$ for a Plummer force law.  Further details
and some analysis of these simulations can be found in Eke, Cole \&
Frenk (\shortcite{ECF96}) and in Cole et al. \shortcite{mock}.

\subsection{Biased Galaxy Catalogues}
\label{sec:bias}

Results from recent galaxy surveys suggest that galaxies do not perfectly 
trace the mass distribution but rather pick out the highly non-linear 
regions.  For instance, Peacock \& Dodds (\shortcite{PD94}) show 
there are in fact different bias factors for differently 
selected galaxy samples, eg. IRAS galaxies and radio galaxies.  
Many of these recent investigations favour a value of $\beta \approx 0.5$, 
so for a flat universe with zero cosmological constant a significant 
amount of biasing is needed to match observations.  To this end 
we employ three different methods of biasing our $\Omega_0=1.0$ simulation, 
in order to find out how robust our parameter estimation techniques are to 
the precise bias prescription.  

In each method we build up a `probability field' on a grid, based on 
the density field smoothed on a particular scale.  We then randomly sample 
the particles using the values of the probability field at their nearest 
grid points as their Poisson mean probability of being selected as galaxies.

The first two methods we employ are Lagrangian, in the sense 
that the selection probability is based on the {\em initial} density 
field.  The third is an Eulerian one, where the galaxy distribution is 
`painted on' depending on the final, evolved density field. 
In each case the bias is chosen to obtain a $\sigma_{8}^{\rm gal}$, 
the variance in galaxy counts in spheres of radius $8\hmpc$, consistent 
with the results of Maddox, Efstathiou \& Sutherland (\shortcite{apmIII}), 
who find $\sigma_{8}^{\rm gal} = 0.96$ for the APM Galaxy Survey.  Thus, the 
bias factor, defined as  $\sigma_{8}^{\rm gal}/\sigma_{8}$, has value 
$b=1.81$.  All our methods are local schemes based on the value of the 
overdensity at a given point.  The methods are expected to produce a 
constant bias on large scales.

\begin{enumerate}
\item[{\bf Method 1}]
High Peaks Biasing.  For this method we smooth the initial density field
by applying a sharp cut-off in the power at a scale $r_{\rm s}$, 
set to be, approximately, the shortest resolvable wavelength in the box, 
$r_{\rm s} \approx 4 \hmpc$.  We then divide by the resulting mass variance, 
$\sigma_{\rm s}$, to obtain a new 
variable, $\nu(\r) = \delta(\r)/\sigma_{\rm s}$.  We use the results 
of Bardeen et al. (\shortcite{BBKS}) to calculate the mean number of 
peaks above a certain threshold, $\nu_{\tau}$, for a random field, 
Gaussian-smoothed on a scale $r_{\rm gal} = 0.54\hmpc$, ie. the number of 
galaxy-scale peaks expected as a function of $\nu$ (cf. \cite{wfde87}).
The value of the threshold is adjusted until the required 
level of bias is obtained, from $\sigma_8^{\rm gal} = b 
\sigma_8$.  We find that $\nu_{\tau} = 0.99$ gives the correct normalization.

\item[{\bf Method 2}]
For this method we obtain $\nu(\r)$ by smoothing the initial density 
with a Gaussian of scale $r_{\rm s} = 5\hmpc$.  Our biasing 
probability field is then a continuous function of $\nu$ and two 
adjustable parameters, $\alpha$ and $\beta$, given by
\begin{equation}
P(\nu) = A \, \exp(\alpha\nu+\beta\nu \mod{\nu})
\end{equation}
where the normalization constant, $A$, is determined by the constraint
\begin{equation}
	\int P(\nu) \ \frac{\exp(-\nu^2/2)}{\sqrt{2 \pi}} \d{\nu} = 1 .
\end{equation}
In the limit of small ($\delta \ll 1$) large-scale perturbations in the 
initial density field, it can be shown that the bias on large scales is 
given by 
\begin{equation}
  b =  1 + \frac{\alpha}{\sigma_{\rm s}}
+\frac{2\beta}{\sigma_{\rm s}}\, \int P(\nu) \,\mod{\nu} \
\frac{\exp(-\nu^2/2)}{\sqrt{2 \pi} } \d{\nu} .
\end{equation}
To fix $\alpha$ and $\beta$ we demand that the rms galaxy density
fluctuations in two different sized cubic cells ($5$ and $20\hmpc$) 
match those predicted by assuming a power spectrum shape from the 
result for the APM survey found by 
Baugh \& Gazta\~naga (\shortcite{baugh96}).   
\item[{\bf Method 3}]

In this model the final density field is smoothed on a scale 
$r_{s} = 5\hmpc$.  The selection probability field has a value
0 or 1 according to whether the overdensity is below or above a 
threshold, $\tau$.  We find that $\tau=1.53$ provides the 
required bias.  The particle distribution resulting from this scheme 
is expected to contain large voids since the sharp cut-off at 
the threshold is extremely effective at evacuating underdense regions 
(\cite{weincole}).   
\end{enumerate}

It is hoped that the three methods described here span a broad range of 
plausible biasing prescriptions that might occur in the real 
universe, and that, if we can develop techniques that are unaffected by 
which of these schemes is chosen, these techniques will be insensitive to 
the way in which bias comes about.  Further details about biasing schemes 
will follow in Cole et al. \shortcite{mock}.  

The density of the biased catalogues is chosen to be one galaxy per 
$25 (\hmpc)^3$, 
roughly four times $\phi_{*}$, the observed number density of $L_{*}$
galaxies (eg. \cite[\S 5]{PJE93};\cite{dukstII}).  
This high sampling density has 
been adopted as we are interested in investigating systematic effects 
on the redshift-space distortions, and so we want to minimize the 
statistical uncertainties produced by sparse sampling.

\subsection{Zel'dovich approximation simulations}
\label{sec:zas}
We attempt to model the redshift-space distortions in our \nbody 
simulations using a Monte Carlo implementation of the Zel'dovich 
approximation.  Our Monte Carlo realizations have the same $192^3$ 
particle grid and initial density field parameters as the corresponding 
\nbody simulations.  
Exactly the same method is used to perturb the particles 
as that employed to get the initial displacements in the 
\nbody simulation, only with much larger perturbation amplitude 
corresponding to $z=0$ rather than the starting redshift of the 
simulation. For the \nbody simulations we populate all the
modes in the $192^3$ $k$-space grid including those in the
corners of the cube. However for these realizations of the
ZA, we make our treatment consistent with the analytic calculation
that we outline in \secref{ZA}, by smoothing the density field in the 
same way, ie. by truncating the power spectrum isotropically at 
$k_{\rm Nq} = 1.743\himpc$. 

These realizations of the ZA can be biased using exactly the same 
prescriptions that we use for the \nbody simulations.
The catalogues thus created are then subjected to the 
same multipole analysis as our \nbody simulations.

\subsection{Estimation of multipole moments}
We ensure that the distant observer approximation is satisfied 
by assigning redshift-space positions to galaxies based on their 
distance along the $x$-axis of the simulation.  Effectively the 
simulation box is placed at infinity and the line of sight 
aligned with the $x$-axis. To minimise the noise in our estimates
we repeat the analysis with the simulation rotated to align the
line of sight with the $y$- and then $z$-axes and average the results.

The galaxies in the simulation are assigned to a $192^{3}$ 
grid using the cloud-in-cell (CIC) weighting scheme
(\cite{edwf85}).  The density 
field thus produced then has the average galaxy density 
subtracted from it and is divided by this average density.  
The resultant grid is thus the fractional overdensity field, $\delta(\r)$, 
and the grid is fast Fourier transformed to get 
$\delta(\k)$.  The effect of the CIC assignment is to doubly convolve 
the density field with the shape of a grid cell.  Thus to deconvolve 
we divide by the square of Fourier transform of the CIC window function.  
The power spectrum estimator is then $\delta(\k)\delta(\k)^{*}$.

For the biased galaxy simulations a constant equal to the inverse 
of the number density of galaxies in the catalogue is subtracted at this 
point to account for the shot noise introduced by the Poisson 
sampling of the galaxies.  This is not necessary in the unbiased 
cases as here we use all the particles rather than a Poisson sample.

We now divide $k$-space in spherical shells of width 
$\Delta k = 2\pi/L_{\rm box}$. For each shell we estimate
the multipole moments of the power spectrum by approximating the
integral of $\mu$ in \eqnref{multipol} by a discrete sum over modes.
This method works perfectly well at high-$k$, but for the first few
shells where the sampling of $k$-space results in poor sampling in 
$\mu$ it leads to a small systematic error in $P_2(k)/P_0(k)$.
An alternative method of estimating $P_2(k)$ and $P_0(k)$, which
can readily be employed for low-$k$, is to make a least squares
fit to the $\mu$ dependence of $P(k,\mu)$ using a basis of the
Legendre polynomials $L_0$,$L_2$ and $L_4$. This second method
produces a less noisy estimator but still suffers from a small
systematic error due to the poor sampling in $\mu$.
We therefore apply a small ($5$\% for the lowest wavenumber), 
empirically derived correction to our first five bins in $k$ to 
compensate for this. 

The lowest $k$-mode probed using this method is 
$k=({2\pi}/{L_{\rm box}})$, ie. when one wavelength fills 
the whole box.  The highest is the Nyquist
frequency, $k_{\rm Nq}=({2\pi}/{L_{\rm box}})({N}/{2})$, where $N$ is 
the number of cells along one side of the simulation box.  
We do not expect our treatment to work all the way up to 
the Nyquist frequency, since the deconvolution correction can never 
recover all the information lost by assigning particles to a discrete grid.  
This effect can be seen in \figref{p0} 
where the power spectrum turns up for the very highest $k$-values. 
For this reason we limit our 
analysis rather conservatively to modes with $k < \frac{1}{2} k_{\rm Nq}$.

\section{Analytic Models}
\label{sec:mods}

\begin{figure*}
\centering
\centerline{\epsfxsize = 16.0 cm \epsfbox{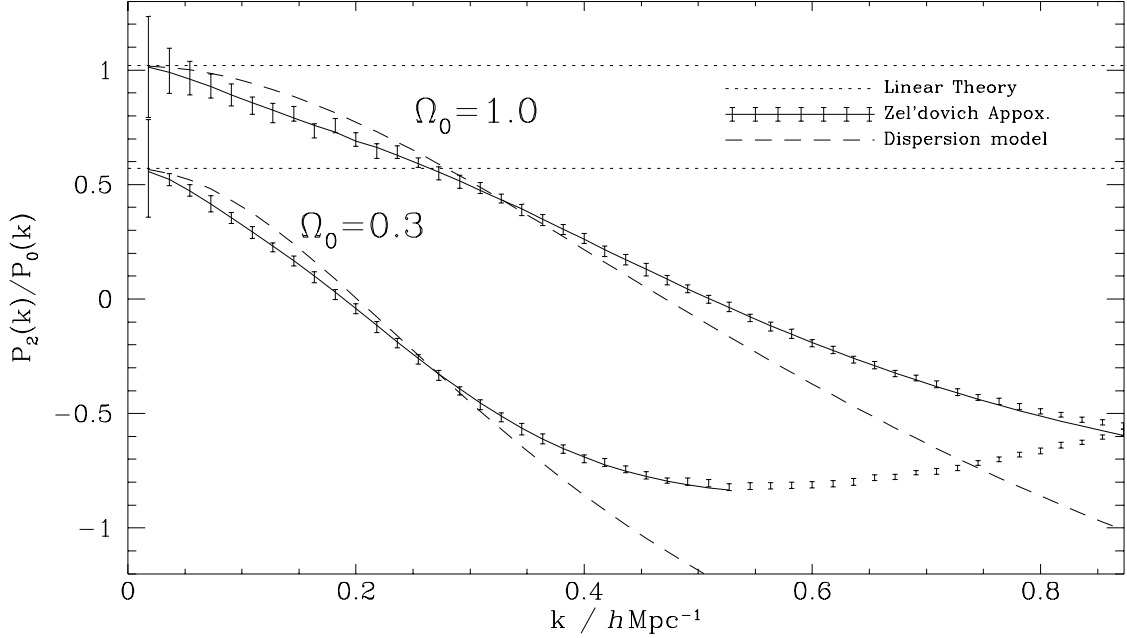}}
\caption{Model predictions for the redshift space 
power spectrum quadrupole to monopole ratio $P_2(k)/P_0(k)$. 
The two sets of curves are for two unbiased ($b=1$)
CDM models with $\Gamma=0.25$ power spectra.
The upper curves are for 
$\Omega_0=1.0$, $\sigma_8=0.55$, 
while the lower model is $\Omega_0=0.3, \Lambda_0=0.7$ and 
$\sigma_8=1.13$.  
In both cases the curves show the constant ratio predicted by 
linear theory (dotted), the analytic Zel'dovich Approximation result 
(solid), and examples of the dispersion model
(dashed) with $\sigma_v=300 \kms$ for $\Omega_0=1.0$ and 
$\sigma_v=500 \kms$ for $\Omega_0=0.3$.
The error bars show the result of averaging
forty random  Monte Carlo realizations of the Zel'dovich Approximation.
These error bars indicate the 
standard deviation of the sample, and thus are typical of the error
expected in a single realization.}
\label{fig:q1}
\end{figure*}

If it is assumed that the velocity field, $\v(\r)$, is generated via 
gravitational instability, then it can be related to the underlying
mass density field. Here we consider three analytic models 
for the relationship of the velocity field to the density field,
which lead to quantitative descriptions of the redshift-space distortion
of galaxy clustering.  The first is base purely on linear theory.  In the 
second, the effect of non-linear velocities in galaxy clusters is modelled 
by adding a random velocity dispersion to the velocity predicted by 
linear theory.  The magnitude of this dispersion, $\sigma_v$, is not 
predicted by the model, but is expected to bear some relation to the 
typical thermal velocity of galaxies within groups and clusters.  The 
third model is the Zel'dovich approximation (ZA) (\cite{ZA}).  This 
model provides a complete description of the evolved density and 
velocity fields, which is expected to be accurate in the quasi-linear 
regime where the overdensity is of order unity.  Thus this model has 
the potential to give an accurate physical description of redshift-space 
clustering.  These three models are described in detail below and 
compared in \figref{q1}.
In each case we assume that the distant
observer approximation is well satisfied and that we are dealing
with volume limited galaxy samples.  The effects of relaxing these
assumptions are discussed in Kaiser (\shortcite{NK}).

\subsection{Linear Theory}

In the regime of linear theory Kaiser (\shortcite{NK})
showed that the redshift-space power spectrum is related 
to the real space power spectrum in a very simple way 
\begin{equation}
\label{eqn:lin}
P(k,\mu) = P(k) (1 + \beta \mu^2 )^2 , \quad
\beta \equiv f(\Omega_0)/b \approx \Omega_0^{0.6}/b.
\end{equation}
The function $f(\Omega_0)\approx\Omega_0^{0.6}$ is the
logarithmic derivative of the fluctuation growth rate (see
\cite[\S 14]{PJE93}; \cite{bouch95}). The bias factor, $b$,
is an assumed constant relating fluctuations in the galaxy 
density to those in the mass.

Applying \eqnref{multipol} for the $l=0$ and $l=2$ modes results in 
\begin{equation}
\label{eqn:qlin}
\frac{P_2}{P_0} = \frac{4\beta/3 + 4\beta^2/7}{1 + 2\beta/3 + \beta^2/5}.
\end{equation}
Thus linear theory predicts a constant quadrupole-to-monopole ratio, 
$P_2/P_0$, independent of scale. This model is shown by the dotted
lines in \figref{q1}.

Here we highlight some of the steps in this derivation 
so that we can compare its assumptions with those of two other 
proposed models.
In linear theory the velocity and galaxy density fields
are related by the first order continuity equation,
\begin{equation}
\nabla . \v(\r) = -\beta \delta(\r).
\end{equation}
Considering the density field as a sum over Fourier components, 
\begin{equation}
\label{eqn:rho}
\delta(\r) =  \sum  \delta(\k) e^{-i\k.\r} ,
\end{equation}
the resulting linear theory prediction for the velocity field is
\begin{equation}
\label{eqn:v}
\v(\r) = -\beta \sum \frac{i\k}{k^{2}} \delta(\k) e^{-i\k.\r} ,
\end{equation}
where we have assumed a linear bias between the galaxy and
mass density perturbations, $\delta(\r)=b \delta_{\rm mass}(\r)$.
This leads to the following expression for the line-of-sight
velocity gradient,
\begin{equation}
\frac{dU}{dr} = - \beta \mu^2 \delta(\r) .
\end{equation}
The results \eqnref{lin} and \eqnref{qlin} are valid only so long 
as each of the four following constraints is satisfied, so that the 
corresponding form of non-linearity can be ignored
(Cole et al. \shortcite[section~2.2]{CFW94});
\begin{enumerate}
\item[1. {\bf dispersion:}]{The velocity dispersion in 
virialized systems is sufficiently
small that it can be ignored, i.e. $k \sigma_v \ll 1$}
\item[2. {\bf dynamical:}]{The linear relationship between velocity and 
overdensity \eqnref{v} is accurate.}
\item[3. {\bf gradient:}]{Second order terms in the gradient of the 
line-of-sight velocity field can be ignored, i.e.  ${dU}/{dr} \ll 1$}
\item[4. {\bf contrast:}]{Second order terms in the galaxy overdensity can be 
ignored,  $\delta(\r) \ll 1$ }
\end{enumerate}

All these constraints will be satisfied on the very largest scales,
but, depending on the present-day amplitude of galaxy clustering and on
the values of $\Omega_0$ and $b$, we expect some or even all of
them to be violated on scales of less than $100 \hmpc$.

\subsection{Linear Theory Plus Dispersion}

The velocity dispersion of galaxies in galaxy clusters
is typically $800\kms$ (eg. White et al. \shortcite{WEF93}).
Thus, from the the constraint that $k \sigma_v \ll 1$, we would 
expect the linear theory result \eqnref{lin}
to apply only on scales where $k \ll 0.125\himpc$. 
Since we are interested in clustering on scales from $10-200\hmpc$
($0.03\lsim k \lsim 0.6\himpc$), an accurate model will have to take into 
account these non-linear velocities.
  
  On small scales the redshift space correlation function
has been found to be well modelled as a convolution of the
real-space isotropic correlation function with an exponential
probability distribution function for line-of-sight  
velocities (\cite{bean,davpee,FisherIRAS}).  

Park \etal (\shortcite{park}) point out that, since the convolution 
in $r$-space 
is equivalent to a multiplication in $k$-space, the power spectrum 
is multiplied by the square of the Fourier transform of the velocity
distribution function. Peacock \& Dodds
(\shortcite{PD94}) show that the effects of linear clustering  
and this model of small-scale velocity dispersion can be combined 
to give a redshift space power spectrum
\begin{equation}
\label{eqn:disp}
P(k,\mu) = P(k) (1 + \beta \mu^2 )^2 
\left ( 1 + \frac{(k \mu \sigma_v)^2}{2} \right )^{-2} . 
\end{equation}
For this expression it is considerably harder to perform the integration 
over $\mu$ required in \eqnref{multipol} than for the linear theory case, 
but an analytic expression for $P_2(k)/P_0(k)$ can easily be obtained
with the aid of mathematics packages such as Maple.  
The rather lengthy formulae are given in Cole et al. (\shortcite{CFW95}) 
section~2.1.

This model extends linear theory by relaxing the dispersion 
constraint listed in the previous section, but still implicitly 
assumes the other three.  
The model of the small-scale velocity  dispersion is also simplistic 
in that it takes no account of the fact that the velocity 
dispersion is correlated with the density field, ie. the dispersion 
is higher in high density regions such as galaxy clusters. This means
that the value of $\sigma_v$ used in this model is only an {\em effective}
velocity dispersion which depends on how galaxies populate the clusters,
and thus on the bias parameter, $b$.
Examples of this model are shown by the short dashed curves in 
Fig.~\ref{fig:q1}.

\subsection{The Zel'dovich Approximation}
\label{sec:ZA}

The linear theory approach to modelling the growth of density 
perturbations is only valid if $\delta(\r) \ll 1$.
The refinement to linear theory proposed by Zel'dovich (\shortcite{ZA}) was 
to formulate a Lagrangian approach.  Here each particle is displaced 
from its original position along a straight line defined by the 
direction of the initial velocity field.  In comoving co-ordinates $\r$, 
the final position, is related to ${\bf q}$, the initial position, by
\begin{equation}
\r = {\bf q} + {\bf d}({\bf q},t),  \quad 
{\bf d}({\bf q},t) = \frac{\v({\bf q})}{f(\Omega_0)}.
\label{eqn:za}
\end{equation}

The Zel'dovich Approximation (ZA) is expected to break down at the stage 
when shell-crossing occurs ($\delta(\r) \sim 1)$.  This is because, 
under this model, the particles pass right through caustics as they continue 
to move in the direction of their original velocity.  In contrast, in
\nbody simulations non-linear effects cause the particles to behave as if 
they were `sticky', and galaxies congregate in high density shells
or walls.
If the power spectrum is not truncated or filtered at high spatial 
frequencies, shell-crossing occurs on small scales, with the result that 
small-scale and some large-scale power is erased, and the degree of 
anisotropy is lower than expected.  
We choose to smooth the initial density field for the Zel'dovich realizations 
on small scales by applying 
a sharp cut-off to the power spectrum at $k_{\rm T}=1.743\himpc$,
which corresponds to the Nyquist frequency of the grid used in the \nbody 
simulations.

Fisher \& Nusser (\shortcite{FN96}, henceforth FN96) 
and Taylor \& Hamilton (\shortcite{HT96}) 
obtained an analytic result for the redshift-space distortion produced 
with the ZA.  Following their work we have developed an integration technique 
using an Euler method (\cite{NUMREC}) to deal with diverging oscillations of 
the integrand.  Over the range of $k$ we are interested in, the
method is quite stable, and avoids the complexity 
described in the appendix of Taylor \& Hamilton (\shortcite{HT96}). 
However, our technique breaks down at a scale $k \gsim 0.6/\sigma_8$.
At this point oscillations in the radial integrand become too rapid and of too 
great an amplitude for the method to cope.

This analytic calculation of the quadrupole ratio $P_2(k)/P_0(k)$ is
compared with the results of averaging many Monte Carlo realizations
of the ZA in Fig.~\ref{fig:q1}. The close agreement of the two methods
is very reassuring. It not only demonstrates the accuracy of our numerical
integration technique, but also our implementation of realizations of the 
ZA and the entire procedure of estimating $P_2(k)/P_0(k)$ from particle 
distributions. 

In Fig.~\ref{fig:q1} the $P_2/P_0$ ratio of the $\Omega_0=0.3$ model
reaches a minimum at $k \approx 0.5 \himpc$ and then slowly increases.
The ratio behaves in a similar manner for $\Omega_0=1.0$, but with
a minimum at $k \approx 0.95 \himpc$, which lies off the right-hand side of 
Fig.~\ref{fig:q1}. We attribute this behaviour to the
breakdown of the ZA at scales on which substantial shell crossing 
has occurred.  This occurs on larger scales in the $\Omega_0=0.3$ model
as it has the larger amplitude of density fluctuations with 
$\sigma_8=1.13$ compared with  $\sigma_8= 0.55$ in the $\Omega_0=1.0$ model.
In reality one expects the ratio $P_2(k)/P_0(k)$ to become increasingly
negative on small scales due to the fingers-of-god produced by 
the random velocities of galaxies in virialized groups and clusters.

In our implementation of the ZA we have applied very little smoothing
of the initial density field. We have simply truncated the input power
spectrum at the Nyquist frequency of our particle grid used in the
Monte Carlo realizations described in \secref{zas} to 
facilitate inter-comparison. However, 
Melott, Pellman \& Shandarin (\shortcite{mel}) have shown that, at least 
for some statistics, the correspondence between the ZA and N-body
simulations can be improved by more severe smoothing of the initial
density field. They define a non-linear wavenumber $k_{\rm nl}$ by  
\begin{equation}
\label{eqn:knl}
4\pi \int_0^{k_{\rm nl}}P(k) k^2 \d{k} = 1
\end{equation}
and then smooth the initial density field with a Gaussian
window, $\exp{(-k^2/2k^2_{\rm s})}$, with $k_{\rm s} = p k_{\rm nl}$.
They find that the best choice of the parameter $p$  
depends only weakly on the shape of the power spectrum
and is $p \simeq 1.25$. When comparing to the \nbody simulations
in \secref{anal} we investigate whether this procedure improves the
redshift space predictions of the ZA.

The ZA has several advantages over pure linear theory.
Firstly, equation \eqnref{v} is modified such that the velocity $\v(\r)$
is given in terms of the density field at the initial position ${\bf q}$,
ie. $\r$ is replaced by ${\bf q}$ in \eqnref{rho},
and on the right-hand-side of  \eqnref{v}. This dynamical relation
remains quite accurate until shell crossing occurs, whereas the
linear theory relation is only accurate for $\delta(\r) \ll 1$.
In contrast to the linear theory derivation, 
no further assumptions regarding either the amplitude of
the velocity gradient nor the density fluctuations are made in deriving
the resulting power spectrum. Thus in this respect it is able to deal with
both gradient and contrast non-linearity.
For modelling redshift-space distortions the main shortcoming of the ZA 
is that it does not model the random velocities produced in non-linear relaxed 
structures --
instead it produces its own velocity dispersion on small scales due 
to the shell crossing. Thus in this respect it does not deal with the
dispersion non-linearity.
Also in the form described above it does not explicitly deal with the
effects of biasing. Fisher \& Nusser (\shortcite{FN96}) claim that
to a good approximation bias can be modelled by simply replacing
$f(\Omega_0)$ with $\beta= f(\Omega_0)/b$ -- we investigate this in 
\secref{anal}. 

\subsection{Fitting the models}
Our objective is to illustrate the systematic differences
between the \nbody results and the models rather than to propose 
a methodology to estimate model parameters from real estimates of 
the redshift-space distortion. 
In order to analyze the data from our simulations on quasi-linear scales,
we will perform $\chi^2$ fits for the dispersion and ZA models to the 
\nbody points and restrict the range of $k$ to the regime over which 
each model can reasonably match the data.

The dispersion model provides a functional form for the 
shape of the quadrupole estimator, but says nothing about 
the clustering itself.  Thus, in fitting to the data, no 
assumptions need be made about the shape or amplitude 
of the underlying power spectrum, we simply do a $\chi^2$ 
minimization using $\beta$ and $\sigma_v$ as fitting parameters.

In contrast, the ZA is much more physically motivated 
in that it attempts to model both the clustering and the 
associated velocity field from first principles.
We first attempt to utilize a simple fitting formula for the shape of
$P_2(k)/P_0(k)$ presented in FN96.  This formula depends on $\beta$ 
and the zero-crossing of the quadrupole, $k_{\rm nl}$.  We find that 
using this approximation introduces a significant offset between 
the best-fit and true values of $\beta$.  We therefore opt to
evaluate the full expression derived by FN96 for the expected quadrupole 
to monopole ratio at each stage in making the 
minimum-$\chi^2$ fit.  This expression gives the distortion as a 
function of $\beta$ and the power spectrum shape and normalization 
($\Gamma$ and $\sigma_8$).  Here we simply adopt the values of 
$\Gamma$ and $\sigma_8$ used in the simulations 
and use $\beta$ as the single free parameter in the model.  
For real observations it is hoped that one could simultaneously fit 
for these parameters using estimates of the redshift-space power 
spectrum, $P_0(k)$.

\section{Results}
\label{sec:anal}
\begin{figure*}
\centering
\centerline{\epsfxsize = 16.0 cm  \epsfbox{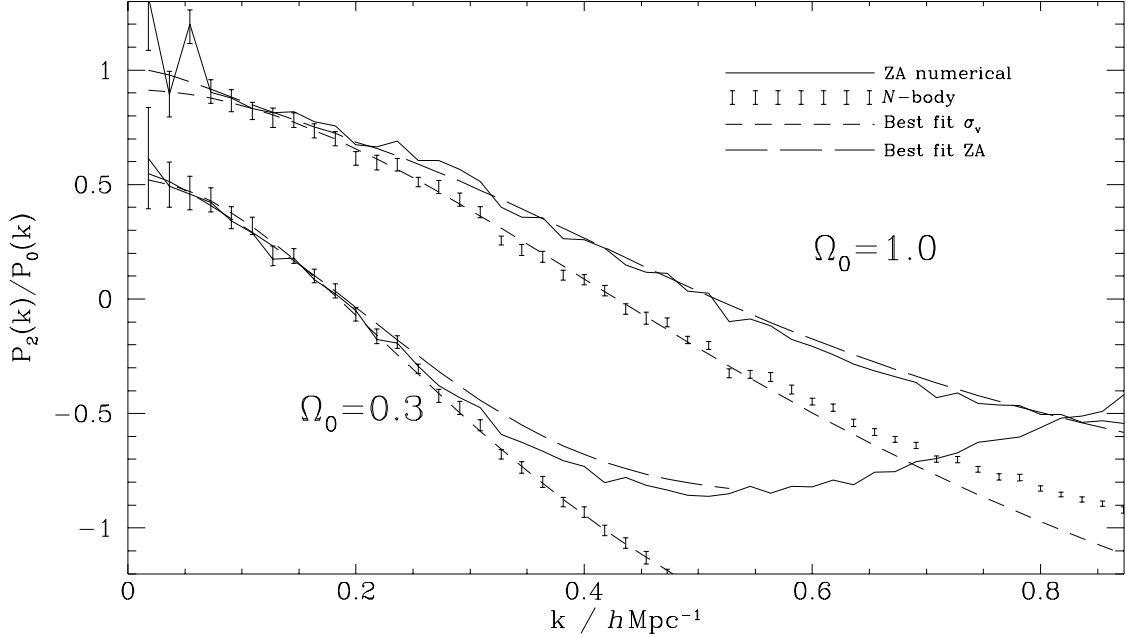}}
\caption{The quadrupole ratio $P_2(k)/P_0(k)$ for two unbiased
\nbody simulations compared with the corresponding Zel'dovich approximation
simulations and two model fits. The upper curves are for $\Omega_0=1$
and $\sigma_8=0.55$ and the lower curves for $\Omega_0=0.3$, $\Lambda_0=0.7$
and $\sigma_8=1.13$. The error bars on the \nbody results are indicative
of the statistical error in our estimates and are estimated from the 
standard deviation between forty ZA realizations. The solid lines show
the results from single ZA realizations starting from the same density 
fields as the \nbody simulations.  The long dashed and short dashed curves
are model fits for the ZA and dispersion model respectively. Only the first
ten points have been used in constraining the ZA fits, and the first thirty 
for constraining the dispersion model.  The corresponding
best fit parameters and confidence intervals are shown \figref{confid}.
}
\label{fig:res1nobias}
\end{figure*}

\begin{figure*}
\centering
\centerline{\epsfxsize = 16.0 cm  \epsfbox{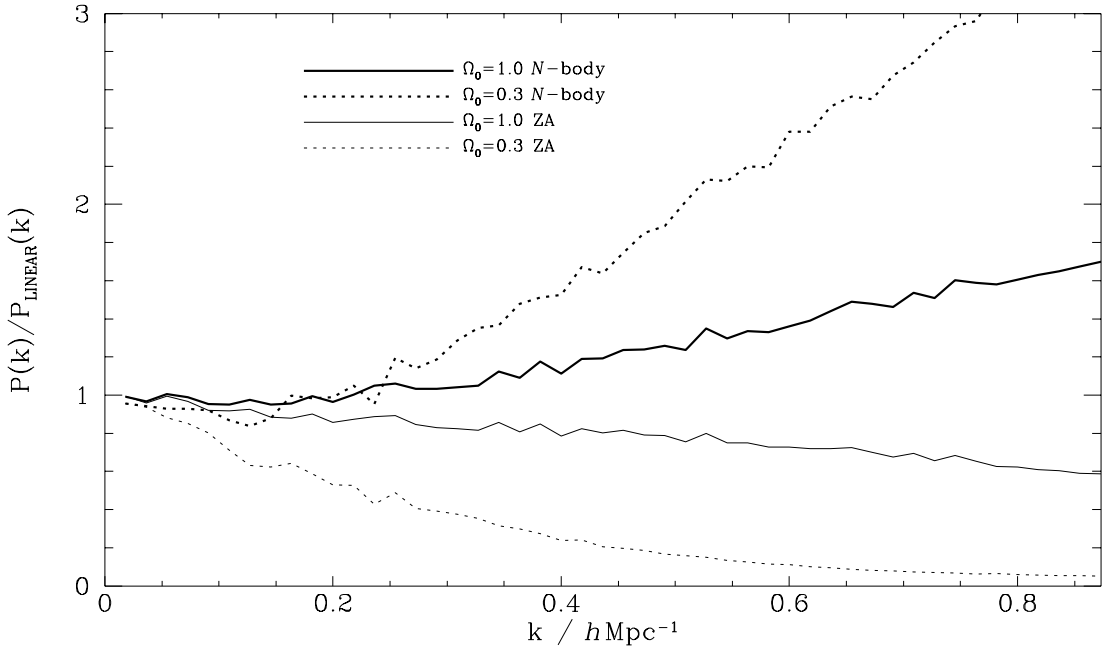}}
\caption{The ratio of estimates of the real-space evolved power spectra,
$P(k)$, to the linear theory predictions, $P_{\rm LINEAR}(k)$,
 for the \nbody simulations and the corresponding ZA simulations.
The models are the same $\Omega_0=1$ and $\Omega_0=0.3$ models studied
in \figref{res1nobias}.
}
\label{fig:mono}
\end{figure*}

\begin{figure*}
\centering
\centerline{\epsfxsize = 16.0 cm  \epsfbox{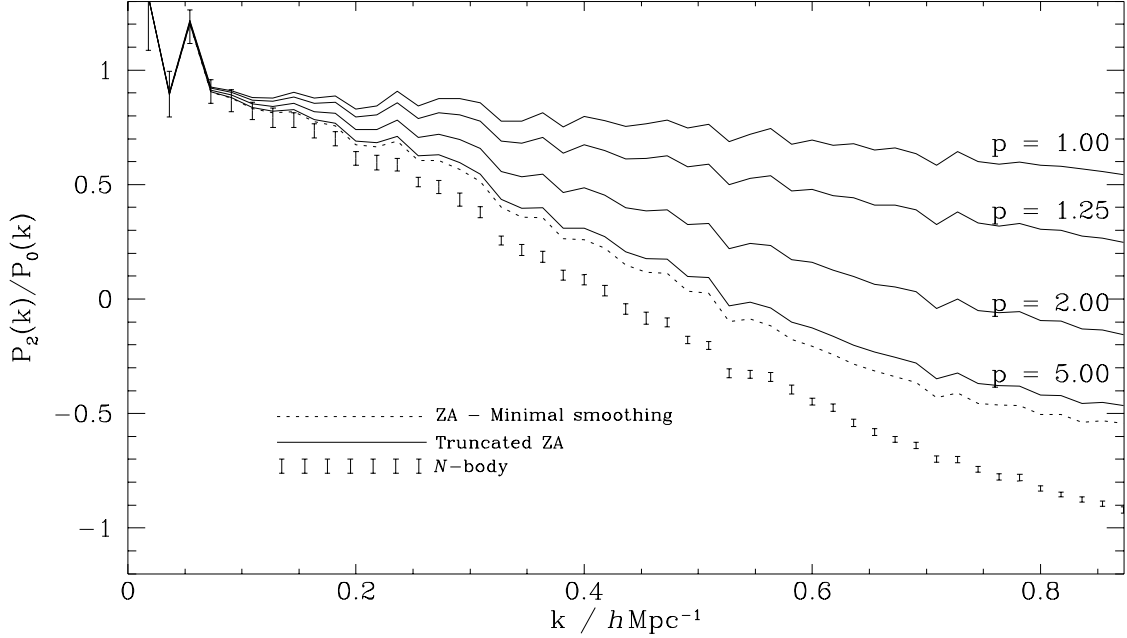}}
\caption{The dependence of the Zel'dovich approximation prediction
for the ratio $P_2(k)/P_0(k)$ on the extent to which the initial density
field is smoothed before applying the ZA. The error bars 
show the \nbody result for the $\Omega_0=1.0$ model taken from 
\figref{res1nobias}. The dotted curve shows the ZA result for the
minimal amount of smoothing, ie, truncation at the Nyquist frequency.
The remaining curves show the effect of gradually increasing the smoothing
using an additional Gaussian smoothing with 
$k_{\rm s} = p k_{\rm nl}$  (see equation \ref{eqn:knl}) and $p=5.0$,$2.0$,$1.25$
and~$1.0$. }
\label{fig:melott}
\end{figure*}

\begin{figure*}
\centering
\centerline{\epsfxsize = 16.0 cm  \epsfbox{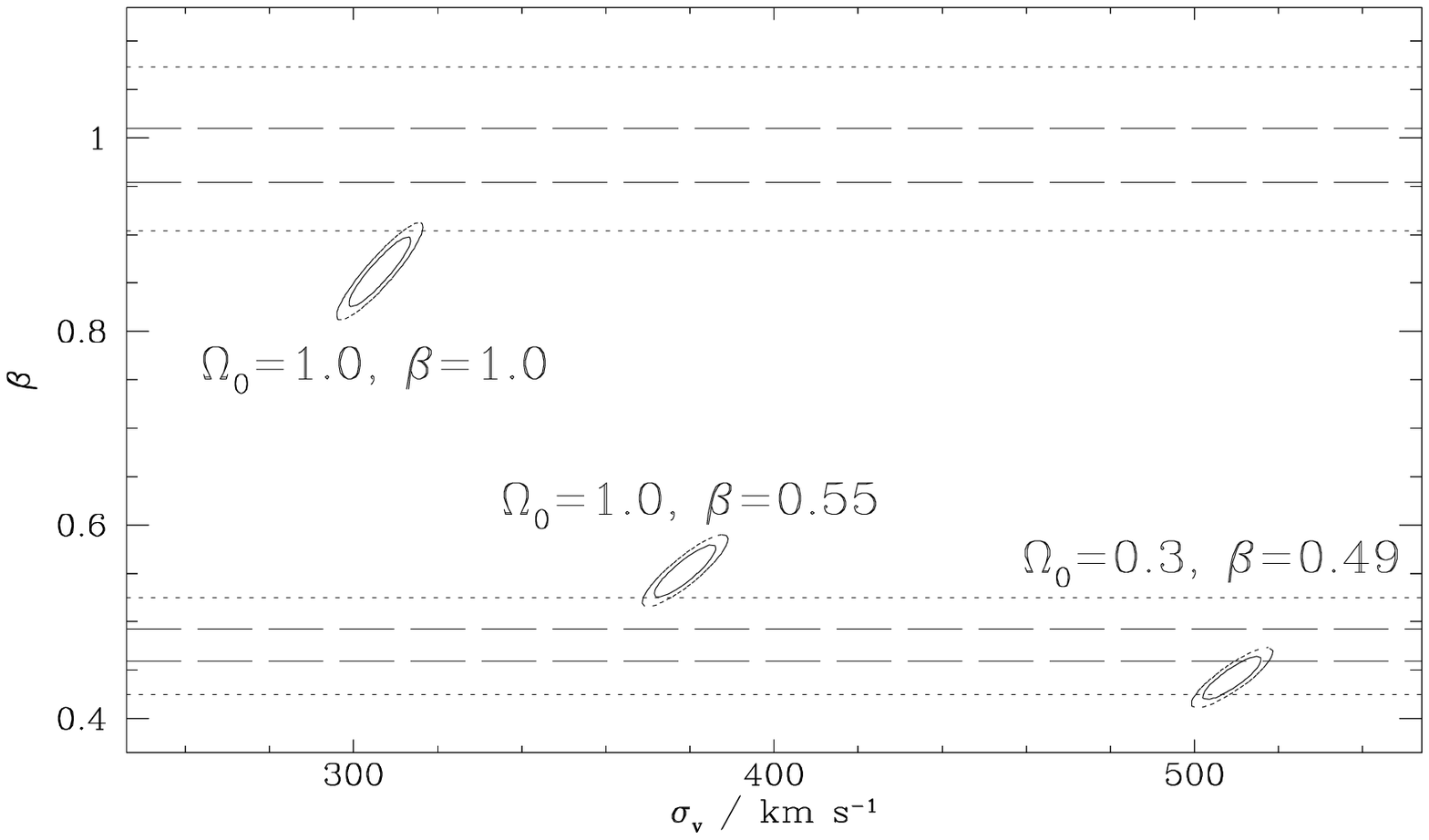}}
\caption{The 68\% and 99.7\% confidence intervals (1- \& 3-$\sigma$)
of the model parameters for the fits shown in \figref{res1nobias} and 
\figref{res2bias}.  
The ellipses are for the two-parameter 
dispersion model for the three different simulations, and the horizontal 
lines for single-parameter ZA model for the two unbiased models for 
which this provides a reasonably good fit.
The best fit values of $\beta$ and the $\chi^2$ per degree of freedom 
for the best fit models are given in \tabref{fits}.}
\label{fig:confid}
\end{figure*}
\subsection{Unbiased Models}
\label{sec:unbiased}

In \figref{res1nobias} we compare the model predictions for the
quadrupole ratio, $P_2(k)/P_0(k)$, with the results of two unbiased
\nbody simulations. The points with error bars show the \nbody results.
The model that produces the lower quadrupole ratios
has $\Omega_0=0.3$, $\Lambda_0=0.7$ and $\sigma_8=1.13$, and that with
the stronger quadrupole signal has $\Omega_0=1$ and $\sigma_8=0.55$.
The error bars placed on the \nbody results are indicative of the
statistical error we expect on an estimate of the quadrupole ratio,
from a single realization, and are obtained by taking the standard 
deviation of forty Monte Carlo realizations of the ZA. It is clear
that, apart from the first few points with $k \lsim 0.07 \himpc$, 
the deviation from the constant ratio predicted by linear theory
is large.

The solid lines in \figref{res1nobias} are each 
for one ZA realization starting from the same initial
density field as the corresponding \nbody simulation. This model
reproduces the \nbody results extremely accurately at low-$k$.
For the $\Omega_0=0.3$ case the ZA and \nbody results match well
up to $k \approx 0.25\himpc$ which is beyond the scale at which
$P_2(k)/P_0(k)$ becomes negative. This result is in complete agreement
with the findings of FN96. The only \nbody results they studied
were for a similarly normalized $\Omega_0=0.3$,$\Lambda_0=0.7$ model. 
In the case of $\Omega_0=1$ we find much less impressive agreement
between the ZA and \nbody simulations. The results begin to diverge
at $k \gsim 0.15 \himpc$ and differ very significantly at $k = 0.425 \himpc$, 
where $P_2(k)/P_0(k)=0$ for the \nbody simulation.

In \figref{mono} we show another comparison between these \nbody
and ZA realizations. Here we show the ratio of estimates of their
final power spectra to the initial power spectra evolved to $z=0$
assuming linear theory. We see that in the \nbody
simulations the non-linear growth of structure leads to an increase
in power on small scales. This is strongest for $\Omega_0=0.3$ 
which has the highest $\sigma_8$ and is therefore the most non-linear.
In contrast the behaviour of the ZA simulations is the exact opposite --
shell crossing erases power on small scales. In this respect the agreement
between ZA and \nbody is particularly poor for the $\Omega_0=0.3$ model.
Even at mildly non-linear scales this discrepancy is 
quite severe, which suggests to us that the success of the ZA at modelling the
ratio $P_2(k)/P_0(k)$ should be treated with skepticism.

The possibility of improving the correspondence with the \nbody results by
smoothing the initial density field before applying the 
ZA, as in the truncated ZA of Melott et al. (\shortcite{mel}), 
is investigated in \figref{melott}.
The introduction of additional smoothing worsens the  agreement between
the \nbody results and those of the ZA. In particular for the degree of
smoothing advocated by Melott et al., corresponding to $p \approx 1.25$,
the agreement between \nbody and ZA is very poor.
Presumably the smoothing being applied is washing 
out the distortions we are trying to measure.  However, we find 
that in real space 
a mild amount of smoothing can produce a power spectrum that is 
slightly closer to the \nbody result, though this is a rather small effect.

The ZA and dispersion model fits are shown by the smooth 
long and short dashed curves in \figref{res1nobias}. The corresponding
best fit model parameters and confidence intervals are shown in 
\figref{confid} and \tabref{fits}. Here the horizontal lines bracket the
68 per cent and 99.7 per cent (1- and 3-$\sigma$) confidence intervals 
of $\beta$  for the 
ZA fit.  The ellipses show the corresponding confidence intervals for
the two-parameter dispersion model.  Both ZA fits give a best-fit 
$\beta$ that is within $1-\sigma$ of the theoretical value, so there 
seems to be no appreciable systematic error in our estimate.  However, 
this result is achieved by limiting the range of the fit to the first 
ten bins in $k$-space, ie. $k < 0.182\himpc$, in both models.  At 
smaller scales than this, particularly in the $\Omega_0=1$ model,
the fits can be seen to strongly overestimate the redshift-space 
distortion.  Thus if a fit were made to data covering a wider range of 
scales, $\beta$ would be severely  underestimated.  
In contrast, the dispersion model fits remain
in reasonable agreement with the data up to much larger $k$.  We 
fit our data up to the thirtieth data bin, corresponding to a length 
scale of $11.5\hmpc$.  For $\Omega_0=0.3$, the dispersion model analysis 
seems to be offset towards a low estimate of $\beta$, but this effect is 
fairly small given the size of the error contours.  The fit itself is a 
very good over the whole range of $k$ in question.  However, for 
$\Omega_0=1$, $\beta$ is systematically underestimated by more than 10 per 
cent at a high level of significance.  This offset can be understood
by reference to the models shown in \figref{q1}.  In comparison to the
ZA, which is an accurate fit at low-$k$, the dispersion model curve
changes slope too rapidly and is thus too shallow at low-$k$.  Thus, 
when data at low-$k$ are used to constrain the fit, the value of $\beta$
is underestimated.  We found this behaviour to be the same for a variant
of the dispersion model in which the small scale random velocity
is assumed to be Gaussian rather than exponentially distributed.
This alternative Gaussian dispersion model produces a worse fit overall,
and a more biased estimate of $\beta$.  The same effect is also expected 
to introduce an offset in the $\Omega_0 =0.3$ simulation, but it is rather smaller 
for this case and so is not easily detected.

As shown in \tabref{fits}, we have calculated the 1-dimensional 
rms peculiar velocity of particles in each simulation, $\sigma_v^{\rm SIM}$, 
in an attempt to look for a physical significance of the $\sigma_v$ value 
found in the dispersion model fit.  It is clear from these measurements that 
there is no simple relationship between the two quantities over the range 
covered by our simulations.

\begin{table}

\vbox {\halign {$\hfil#\hfil$&&\quad$\hfil#\hfil$\cr
\noalign{\hrule\vskip 0.03in}\cr
\noalign{\hrule\vskip 0.08in}
\hbox{$\Omega_0$} & \hbox{$\beta$} & \hbox{model} & \hbox{$\beta_{\rm fit}$} & \hbox{$\chi^2/\nu$} & \hbox{$\sigma_v^{\rm FIT}/\kms$} & \hbox{$\sigma_v^{\rm SIM}/\kms$}  \cr
\noalign{\vskip 0.03in}
\noalign{\hrule\vskip 0.08in}
1.0 & 1.00 & {\rm ZA}   & 0.98 & 2.28 &     &     \cr
1.0 & 1.00 & {\rm disp} & 0.87 & 2.03 & 306 & 381 \cr
0.3 & 0.49 & {\rm ZA}   & 0.48 & 0.86 &     &     \cr
0.3 & 0.49 & {\rm disp} & 0.44 & 0.99 & 509 & 442 \cr
1.0 & 0.55 & {\rm disp} & 0.56 & 1.39 & 379 & 427 \cr
\noalign{\vskip 0.08in}
\noalign{\hrule}
}}
\caption{Best fit $\beta$ and goodness of fit ($\chi^2$ per degree of 
freedom) indicator for our simulations under the Zel'dovich and 
dispersion models.  For the dispersion model we also show the best fit 
values of $\sigma_v$ and compare with the 1-dimensional velocity 
dispersion of the simulations, $\sigma_v^{\rm SIM}$.  Error contours 
for the fits are shown in \figref{confid}.}
\label{tab:fits}
\end{table}

\subsection{Biased Models}

\begin{figure*}
\centering
\centerline{\epsfxsize = 16.0 cm  \epsfbox{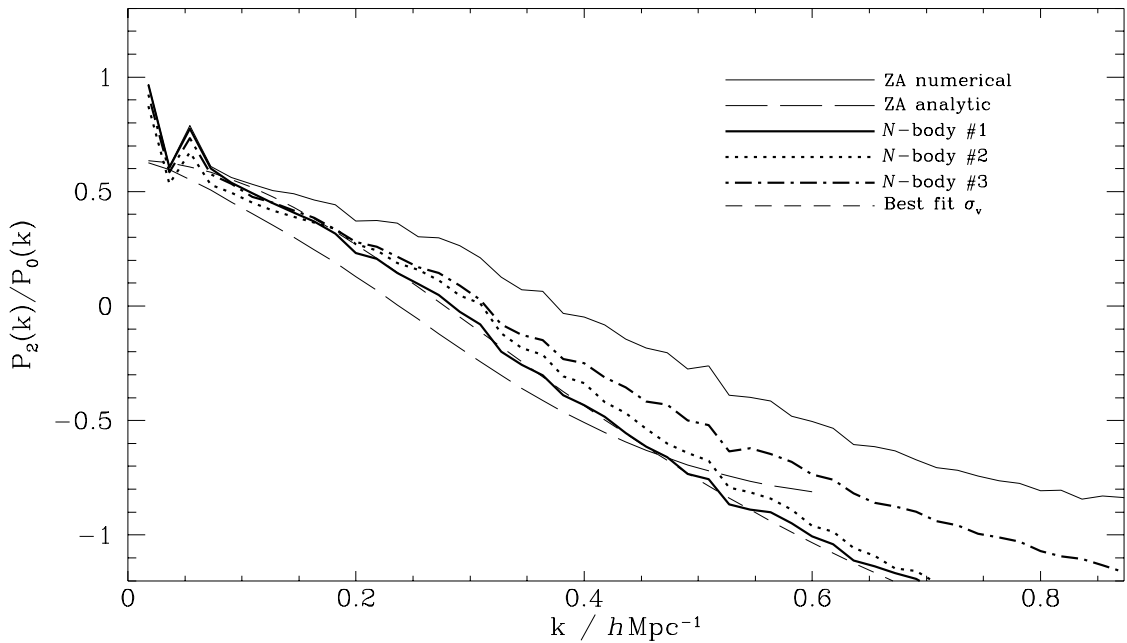}}
\caption{The quadrupole ratio, $P_2(k)/P_0(k)$, for biased $\Omega_0=1$
galaxy catalogues with $b=\sigma_8^{\rm gal}/\sigma_8=1.81$, $\beta=0.55$.
The \nbody lines show the results for our three different methods 
of constructing biased galaxy catalogues from the \nbody simulation, 
as outlined in \secref{bias}.
The thin solid line shows the result of a ZA realization that has been biased
in the same as one of the \nbody catalogues. The long dashed curve
shows the prediction of the ZA made by assuming 
$\sigma_8= \sigma_8^{\rm gal}$ and $\Omega_0^{0.6}=\beta$ as proposed by FN96.
The short dashed line is a fit of the dispersion model to one of the
\nbody catalogues.
}
\label{fig:res2bias}
\end{figure*}

In \figref{res2bias} we show the model predictions for the quadrupole
ratio, $P_2(k)/P_0(k)$, for a biased model with $\Omega_0=1.0$.
The heavy curves show the results of selecting biased galaxy
catalogues by the three methods described in \secref{bias}.
The differences between the models at large scales are purely 
statistical, since different galaxies have been selected in the 
different samples.  At high $k$ the two Lagrangian methods agree 
well, but the Eulerian one systematically shows less distortion.  
This is to be expected as this method populates evenly all regions 
with density higher than the threshold, rather than giving extra 
weight to the areas with higher density (and thus velocity dispersion), 
as the other two models do.

The long dashed curve shows the prediction for the analytic ZA
where, as suggested by FN96, the model adopted assumes that
galaxies are unbiased, $\sigma_8= \sigma_8^{\rm gal}$, but with
$\Omega_0^{0.6}=1/1.81$ to match the value of $\beta=1/b$.
This curve should be compared with the results from a biased
catalogue constructed from a realization of the ZA which is shown 
by the thin solid curve. Clearly these two curves do not agree.
Thus in the ZA, bias cannot be taken account of by merely modifying 
the value of $f(\Omega_0)$ as suggested by FN96. This is perhaps
to be expected as bias will not only boost the galaxy density fluctuations
with respect to the underlying mass distribution, but will also
preferentially place galaxies in dense structures where random
velocities produced by virialized structures are largest.
The small scale velocity field produced by shell crossing in the
ZA is not a good model of the random velocities produced in non-linear
regions. It is for this reason that the biased ZA realization is a
poor fit to the \nbody results. It diverges from the \nbody result
more rapidly than the corresponding model for the unbiased $\Omega_0=1$ 
simulation.

The short dashed curve is a fit of the dispersion model to the
\nbody peaks biased galaxy catalogue.
This model, with an exponential velocity distribution, 
determines $\beta$ very accurately and is a good fit to the data 
over this range of scale.
The fitted velocity dispersion is $\sigma_v = 380 \kms$, 
$70 \kms$ higher than the unbiased version of the same
\nbody simulation, which illustrates how bias preferentially
places galaxies in environments with higher thermal velocities.

\section{Discussion}
\label{sec:conc}

We have used a set of high resolution \nbody simulations to investigate 
the accuracy of two models of the redshift-space distortion of galaxy
clustering on scales of $10$--$200\hmpc$. We conclude that neither
the Zel'dovich Approximation (\cite{FN96,HT96})
nor the linear theory plus random velocity dispersion model 
(\cite{PD94}) provide an accurate description of redshift-space clustering.
In general both models could lead to significant systematic errors
in the estimation of $\beta \equiv \Omega_0^{0.6}/b$ when applied to
the high precision data that will be available from the large 
2dF and SDSS galaxy redshift surveys.

The Zel'dovich Approximation provides a good fit in the unbiased cases on 
the very largest scales that we have investigated.  However, only for
the low $\Omega_0$ model that we examined does this model remain accurate
over the full range of scales for which we measure a positive quadrupole
distortion ($P_2(k)/P_0(k)>0$).  In the case of biased galaxy catalogues
the ZA is a very poor description of redshift clustering apart from on
the very largest scales.  Smoothing the initial density field prior to
applying the ZA, as in the Truncated ZA (\cite{mel}), was found
to further worsen the agreement with the \nbody results of redshift-space
distortions.  In general, then, we would only advocate 
using the ZA for analysis of unbiased universes on scales down to 
$\sim 40\hmpc$.

\begin{figure*}
\centering
\centerline{\epsfxsize = 16.0 cm  \epsfbox{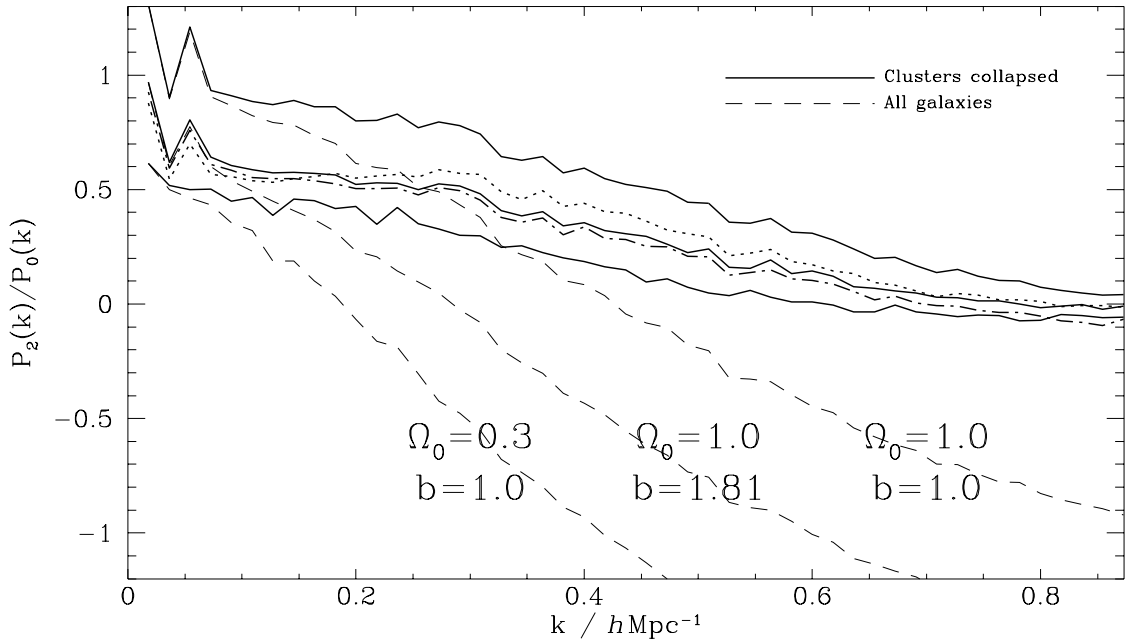}}
\caption{A comparison of the quadrupole ratio, $P_2(k)/P_0(k)$ 
before and after identifying and collapsing the galaxy clusters.  
The dashed lines show the original shape for the three simulations, where 
we have used the peaks biasing method for the biased case.  The solid 
lines the result after cluster collapse for the three models.  We also plot 
the other two variants of the biasing model in the collapsed case, 
using a dotted line for method 2 and dot-dashed for method 3.}

\label{fig:clus}
\end{figure*}

The two-parameter dispersion model was found to produce more acceptable
fits to the \nbody results. However, only in the cases in which
the small scale velocity dispersion was large did these fits yield
values of $\beta$ that were not significantly biased with respect to the
true values. This model also yields a value for the 
velocity dispersion parameter $\sigma_v$, but this depends on the
degree and form of galaxy bias and so is difficult to directly relate
to an interesting physical quantity.  It should be noted that, although 
the dispersion model was unable to deal with the unbiased $\Omega_0=1.0$ 
simulation, and in fact extracted a value of $\beta$ offset from the true 
value by several-sigma, 
results from a variety of cosmological methods are making 
it seem increasingly unlikely that we live in such a universe -- 
constraints from the observed $\sigma_8$ and current $\beta$ estimates 
require a low-$\Omega_0$ or biased cosmology.  Thus the dispersion model 
may well prove to be a useful tool for extracting information from 
redshift surveys on intermediate scales.

There appear to be two distinct ways to proceed to remedy or by-pass
the above short-comings.

1) Improve the analytic models so as to produce accurate predictions
over the full range of scales that will be probed by the redshift surveys.

A promising approach is an extension of the model discussed in
Fisher (\shortcite{KBF}), where linear distortion effects on 
$\xi(\sigma,\pi)$ are modelled as a convolution with a Gaussian 
velocity distribution with scale- and orientation-dependent velocity 
dispersion.  Fisher suggests that this model may be adapted to deal with 
distortions in the highly non-linear regime by replacing the 
Gaussian with a generalized distribution function which is 
Gaussian on large scales but tends to an exponential with isotropic 
and constant velocity dispersion on small scales.

Hamilton et al. (\shortcite{hklm}) developed an accurate method 
of obtaining the non-linear correlation function in real space, 
$\xi(r)$, using a universal scaling relation.  
Mo, Jing, \& B\"orner (\shortcite{MJB96}) outline a technique by 
which the quantities which define the velocity field (ie. mean 
pairwise peculiar velocity, $v_{12}(r)$, pairwise peculiar 
velocity dispersion, $\avg{v_{12}^2(r)}$, and mean 
square peculiar velocity, $\avg{v_1^2}$), can be modelled for 
a given cosmology and initial power spectrum.  

Given reliable models for these quantities, Fisher's treatment 
should provide a way of translating $\xi(r)$ into the redshift space
statistic $\xi(\sigma,\pi)$, and thus calculating precisely the anisotropy 
expected on any given scale.  However, this method is still subject to 
problems -- namely, finding out exactly how the velocity distribution 
function behaves on intermediate scales, and developing a way for 
modelling the effects of galaxy bias.  We intend to investigate this 
in a future paper.

2) As our computers increase in memory and CPU power, the `brute 
force' approach of simply running \nbody simulations for the whole 
range of parameter space becomes increasingly plausible.  
Here the dispersion model may be very useful in 
fitting and suggesting the approximate value of $\beta$, and 
thus putting restrictions on the combinations of parameter values 
worth investigating. 

If one is resorting to simulations then it may be possible to 
filter the data to remove some non-linearities before commencing 
the analysis of redshift-space distortions.  An interesting 
procedure for doing this 
is to identify and collapse galaxy clusters. Since
their high velocity dispersion is responsible for much of the
suppression of $P_2/P_0$, their removal might lead to more robust
estimates of $\beta$.  
\figref{clus} illustrates the effect of collapsing clusters in 
our simulations.  As expected, the anisotropy is much closer to 
the linear theory prediction on large and intermediate scales.  
On small scales the quadrupole ratio converges on zero rather than 
becoming strongly negative.  Cluster membership was 
established using a friends-of-friends algorithm, in which any pair of 
galaxies with separation less than $0.2$ times the mean separation 
are classed as being in the same cluster.  Each cluster member then has 
its velocity replaced with the mean velocity of the 
cluster itself, effectively eliminating the internal velocity 
dispersion of these virialized structures.

The identification of galaxy clusters will be much more complex 
when dealing with real galaxy redshift surveys.  
Moore, Frenk \& White (\shortcite{moore}) describe a way of applying 
this technique to the CfA survey using a redshift- and 
orientation-dependent linking length to define groups.  Whatever 
algorithm one adopts can always be tested by running it on mock 
galaxy catalogues made from the \nbody simulations.

Despite the failings of existing models pointed out in this work, 
redshift-space distortion analysis remains very interesting and promises to 
be of great use in extracting information from the next generation of 
galaxy surveys.  

\section*{Acknowledgements}
The authors wish to thank Carlos Frenk, David Weinberg and Scott 
Croom for some useful criticisms and Karl Fisher for advice 
on evaluating the Zel'dovich approximation integral.  
SC acknowledges the support of 
a PPARC Advanced Fellowship.  SJH acknowledges the support of
a PPARC Research Studentship.


\end{document}